\documentstyle[epsfig,aps,prb]{revtex}
\newcommand\beq{\begin{equation}}
\newcommand\eeq{\end{equation}}
\newcommand\bea{\begin{eqnarray}}
\newcommand\eea{\end{eqnarray}}
\newcommand\ed{E /\delta }
\newcommand\qd{q /\delta }
\newcommand\psr{\psi_R }
\newcommand\psl{\psi_L }
\newcommand\psp{\psi_+ }
\newcommand\psm{\psi_- }

\begin{document}

\begin{center}
{\Large One-dimensional fermions with incommensurate hopping close to
dimerization}
\end{center}

\vskip .5 true cm
\centerline{\bf Diptiman Sen \footnote{E-mail address: 
diptiman@cts.iisc.ernet.in} and Siddhartha Lal \footnote{E-mail address:
sanjayl@cts.iisc.ernet.in}}
\vskip .5 true cm

\centerline{\it Centre for Theoretical Studies, Indian Institute of Science,}  
\centerline{\it Bangalore 560012, India} 
\vskip .5 true cm

\begin{abstract}

We study the spectrum of fermions hopping on a chain with a weak 
incommensuration close to dimerization; both $q$, the deviation of the wave 
number from $\pi$, and $\delta$, the strength of the incommensuration, are 
small. For free fermions, we use a continuum Dirac theory to show that there 
are an infinite number of bands which meet at zero energy 
as $q$ approaches zero. In the limit that the ratio $q/ 
\delta \rightarrow 0$, the number of states lying inside the $q=0$ gap is 
nonzero and equal to $2 \delta /\pi^2$. Thus the limit $q \rightarrow 0$ 
differs from $q=0$; this can be seen clearly in the behavior of the specific 
heat at low temperature. For interacting fermions or the $XXZ$ spin-$1/2$ 
chain, we use bosonization to argue that similar results hold. 

\end{abstract}
\vskip .5 true cm

~~~~~~ PACS number: ~71.10.Fd, ~75.10.Jm
\vskip 1 true cm

%\newpage

One-dimensional lattice models with incommensurate hopping elements or 
on-site potentials have been studied for many years from different points 
of view \cite{sok,luc,val}. Many unusual properties of the quantum
spectra and wave functions have been discovered for various kinds of 
aperiodicity \cite{ost}. Physically, such models have
applications to several problems such as incommensurate crystals 
\cite{lyu} and semiconductor heterojunctions \cite{mer}. If the 
incommensuration is close to dimerization (wave number $\pi$), then the 
models have an additional interest in the context of metal-insulator 
transitions or spin-Peierls compounds \cite{bha,nak}. In this paper, we study 
a model of fermions with a hopping which has a small incommensurate term 
close to dimerization. We will show that this has some unusual properties,
particularly inside the gap which exists {\it exactly} at dimerization. The 
major surprise is that states appear inside the gap {\it as soon as} we move 
away from dimerization. These states carry a finite weight, and 
therefore contribute to quantities like the specific heat at low 
temperatures. The bulk of our discussion is for free fermions or the
$XY$ spin-$1/2$ chain for which quantitatively accurate results can be 
obtained by analytical methods. However, we will argue at the end that the 
same results should hold for interacting fermions or the $XXZ$ spin-$1/2$ 
chain, which is the more general and interesting case.

We begin with the following Hamiltonian for noninteracting and spinless 
fermions on a lattice 
\bea
H ~&=&~ - ~\frac{1}{2} ~\sum_n ~J_n ~(~ c_n^\dagger ~c_{n+1} ~+~ 
c_{n+1}^\dagger c_n ~) ~-~ \mu ~\sum_n ~c_n^\dagger ~c_n ~, \nonumber \\
J_n ~&=&~ 1 ~+~ \delta ~\cos ~(\pi +q) n ~,
\label{ham1}
\eea
where we will assume that $\delta \ll 1$ and $q \ll \pi$. We set the chemical 
potential $\mu =0$ as we are interested in energies close to zero.
If $\delta=0$, the model can be easily solved; the 
dispersion relation is $E(k) = -\cos k$. In the ground state, all the states 
with momenta lying in the range $[-k_F , k_F ]$ are filled, where the Fermi 
momentum $k_F =\pi /2$. The Fermi velocity is equal to $\sin k_F = 1$. 

If $\delta \ne 0$ but $q=0$, i.e., with dimerization, the model can still 
be solved analytically. There is an energy gap extending from $-\delta$ to 
$\delta$. Let us now consider nonzero values of $q$. [Since $\cos (\pi + q)n
= \cos (qn) \cos (\pi n)$, we see that a small value of $q$ is equivalent
to dimerization term whose amplitude $\delta \cos (qn)$ varies slowly with
$n$]. For any rational value of $q/\pi = M/N$, where $M$ and $N$ are 
relatively prime integers, we have a periodic system with period $P$
equal to $N$ if $M+N$ is even and $2N$ is $M+N$ is odd. The one-particle 
spectrum of Eq. (\ref{ham1}) can be found by computing a $2 \times 2$ transfer 
matrix $M(q,E)$ obtained by multiplying together $P$ matrices \cite{ost},
\beq
M (q,E) ~=~ \prod_{n=1}^P ~\Bigl( \begin{array}{cc} 
-2E/J_n & - J_{n-1}/J_n \\
1 & 0 
\end{array} \Bigr) ~.
\eeq
If $\vert {\rm tr} ~M(q,E) \vert \le 2$, the energy $E$ is allowed in the
spectrum; otherwise that energy is not allowed. By following this method and 
sweeping through a large number of values of $q$ and $E$, we obtain the picture
of energy bands and gaps (shaded and unshaded regions, respectively) shown in 
Figs. 1 and 2, taking $\delta =0.05$. (We have sampled the energy values with 
a resolution of $dE = 10^{-6}$). We have scaled both $E$ and $q$ in units 
of $\delta$ because the pictures turn out to depend only on those two 
ratios. We immediately see that the pictures look much more complicated
than the situation in which $q$ is {\it exactly} equal to zero; in that case,
there is precisely one gap extending from $\ed = 0$ to $1$. [We will consider 
only positive values of $q$ and $E$ since it can be shown that the spectrum is 
invariant under either $q \rightarrow -q$ or $E \rightarrow -E$. Thus the
energy bands shown in Figs. 1 and 2 should be understood as continuing to 
negative values of both $E$ and $q$ by reflection about those two axis].

Fig. 1 shows that as $\qd$ increases, the gaps shrink rapidly.
We will show that this can be understood using perturbation theory to $n^{\rm 
th}$ order, where $n$ is an odd integer, and that the gaps shrink as 
$(\delta /q)^n$. We have therefore labeled the gaps in Fig. 1 by the 
integers $n=1,3,5,...$. More interestingly, we observe that all the bands
approach the origin $(q,E)=(0,0)$. The widths of the low-lying bands vanish 
exponentially fast as $\qd \rightarrow 0$ as we will argue from a 
WKB analysis. This makes it impossible to find these portions of the 
bands by looking for energies satisfying $\vert {\rm tr} ~M(q,E) \vert \le 2$
using any reasonable
energy resolution $dE$. We therefore find these thin portions of the
bands by looking for minima of ${\rm tr} M^2 (q,E)$ as functions of 
$E$; these minima yield single points in energy which lie
within a distance of $dE$ of the band. We are thus able to find the 
locations of the thin regions without having to use a resolution smaller than
$dE = 10^{-6}$. In Fig. 2, we show these points (thin regions) which smoothly 
join on to the wider regions of the bands. We also show six curves which 
are the analytical results of a low-energy theory of the model which involves
solving a $1+1$-dimensional Dirac equation in a periodic potential. We will
also prove that the number of bands in the region $0 \le E/\delta \le 1$ 
increases as $2\delta /\pi q$ as $\qd \rightarrow 0$, while the number of 
states in each band is equal to $q/\pi$ when normalized appropriately in the 
thermodynamic limit, i.e., the number of sites $L \rightarrow \infty$. This 
will show that the number of states lying inside the $q=0$ gap is finite 
and equal to $2\delta /\pi^2$ in the limit $q \rightarrow 0$; this 
implies that $q=0$ is a rather singular point.

To begin with, let us understand the gaps for large values of $\qd$ by using 
perturbation theory about $\delta =0$. If $q$ and $\delta$ are both 
small, the states close to zero energy are dominated by momenta near $\pm 
\pi /2$. Since the incommensuration
has Fourier components at momenta $\pm ( \pi + q)$, we see that the gaps above
zero energy result from the breaking of the energy degeneracy between the two 
states with momenta $k_1 = \pi /2 + nq/2$ and $k_{n+1} = -\pi /2 - nq/2$, 
where $n=1,3,5,...$. These two states are connected to each other through 
the $n-1$ successive intermediate states with momenta $k_2 = -\pi /2 + 
(n-2)q/2$, $k_3 = \pi /2 + (n-4)q/2$, $k_4 = -\pi /2 + (n-6)q/2$, ...,
$k_n = \pi /2 - (n-2)q/2$, since we must have $k_{i+1} - k_i = \pm (\pi + q)$
mod $2\pi$. Since all the relevant matrix elements are equal to
$\delta /2$, and the energy denominators are of the order of $q$ each,
we see that the two-fold degeneracy (at the energy $E_1 = E_{n+1} \simeq nq/2$)
gets broken at the $n^{\rm th}$ order
in perturbation theory to produce a gap of the order of $\delta^n /q^{n-1}$.
To be explicit, we find that the upper and lower edges of the two gaps labeled
by $n=1$ and $3$ in Fig. 1 are given by
\bea
\frac{E_{\pm} (1)}{\delta} ~&=&~ \frac{q}{2\delta} ~\pm ~ \frac{1}{2} ~, 
\nonumber \\
\frac{E_{\pm} (3)}{\delta} ~&=&~ \frac{3q}{2\delta} ~+~ \frac{3\delta}{16q} ~
\pm ~ \frac{\delta^2}{32q^2} ~, 
\eea
and we have verified that this agrees well with the numerically obtained 
boundaries. We will show elsewhere that the general formula for the gap 
$\Delta E (n) = E_+ (n) - E_- (n)$ labeled by the odd integer $n=2p+1$ is 
given by
\beq
\frac{\Delta E (n)}{\delta} ~=~ \frac{( \delta /4q )^{2p}}{
( p ! )^2} 
\eeq
to lowest order in $\delta$. 

We can count the number of states in the band lying between gaps $n$ and $n+2$ 
as follows. For $\delta =0$, let us normalize the number of states so that
the total number is $1$; since the momentum $k$ goes from $-\pi$ to $\pi$
in that case, the density of states in momentum space
is $1/2\pi$. If we now turn on a very small value of $\delta$, we find that
the band lying between the gaps $n$ and $n+2$ is made up of
linear combinations of the states lying between the two momenta intervals
$[\pi /2 + nq/2, \pi /2 + (n+2)q/2]$ and $[-\pi /2 - nq/2, -\pi /2 - 
(n+2)q/2]$. The total number of states in these two intervals is $2(q/2\pi) =
q/\pi$. Thus each band contains $q/\pi$ states. Now, this number cannot
change if we change $\delta$, and we can therefore use the same number
below in the opposite limit where $q/\delta$ is small.

Let us examine the more interesting regime where $\qd$ is small
and $\ed < 1$. For analyzing this, it is useful to consider the continuum 
limit. In this limit, the Fermi field $\psi (x) = c_n$ can be written as 
\beq
\psi (x) ~=~ \psr (x) ~\exp ~( i \frac{\pi}{2} x) ~+~ \psl (x) ~\exp ~( - i 
\frac{\pi}{2} x) ~,
\label{fer}
\eeq
where $\psr$ and $\psl$ denote the right- and left-moving fields, respectively;
they vary slowly on the scale of the lattice spacing (which is set equal to 
$1$ here). We substitute (\ref{fer}) in Eq. (\ref{ham1}) and drop terms 
like $\exp (\pm i\pi x)$ which vary rapidly. We then find the following 
Dirac-like equations for the two time-dependent (Heisenberg) fields
\bea
i ~(~\partial /\partial t ~+~ \partial /\partial x ~) ~\psr ~-~ i \delta \cos 
(qx) ~\psl ~&=&~ 0 ~, \nonumber \\
i ~(~\partial /\partial t ~-~ \partial /\partial x ~) ~\psl ~+~ i \delta \cos 
(qx) ~\psr ~&=&~ 0 ~. 
\label{eom1}
\eea
Eq. (\ref{eom1}) can be solved by defining 
$\psi_{\pm} = \psr \pm \psl$ which satisfy the second-order equations
\bea
[ ~ \partial^2 /\partial t^2 ~-~ \partial^2 /\partial x^2 ~+~ \delta^2 
\cos^2 (qx) ~-~ \delta q \sin (qx) ~ ] ~\psp ~&=&~ 0~, \nonumber \\
{[} ~ \partial^2 /\partial t^2 ~-~ \partial^2 /\partial x^2 ~+~ \delta^2 
\cos^2 (qx) ~+~ \delta q \sin (qx) ~ {]} ~\psm ~&=&~ 0~. 
\label{eom2}
\eea
It is sufficient to solve one of these equations, say, for $\psp$, since $\psm$
is related to $\psp$. The energy spectrum can be found by solving the
time-independent equation
\beq
[~-~ \partial^2 /\partial x^2 ~+~ \delta^2 \cos^2 (qx) ~-~ \delta q \sin 
(qx) ~] ~\psp ~=~ E^2 ~\psp ~.
\label{eom3}
\eeq

Eq. (\ref{eom3}) has the form of a Schr\"odinger equation (with ``energy" 
$E^2$) in the presence of a periodic potential. The potential is similar but 
not identical to Mathieu's equation \cite{abr}. We therefore again expect 
bands to form. To simplify the notation, let us shift $x$ by $\pi /2q$ and 
then scale it by a factor of $q$ to make the period equal to $2\pi$. We then 
get
\beq
[~-~ \partial^2 /\partial x^2 ~+~ \frac{\delta^2}{q^2} \sin^2 x ~-~ \frac{
\delta}{q} \cos x ~] ~\psp ~=~ \frac{E^2}{q^2} ~\psp ~.
\eeq
By Floquet's theorem \cite{abr}, the solutions must satisfy $\psp (x+2\pi )=
e^{i\theta} \psp (x)$, where $\theta$ goes from $0$ to $\pm \pi$ from the
bottom of a band to the top. We observe that there is an exact 
zero energy state with $\psp (x) = \exp (~\frac{\delta}{q} \cos x~)$ and
$\psm (x) =0$ for which $\theta =0$. In general, the nonzero energy 
states can only be found numerically.
But if $\qd$ is large, the positions of the low-lying bands (with
$E^2 /q^2 \ll \delta /q$) can be found analytically as follows. For 
energies much lower than the maxima of the potential, we can ignore 
tunneling between the different wells to begin with. Near the bottom of
a single well, we have a simple harmonic potential with small anharmonic
corrections. Ignoring the anharmonic terms, we find the energy levels to be
simply given by
\beq
E_n^2 ~=~ 2 n q \delta ~,
\label{en1}
\eeq
where $n=0,1,2,...$. Next, we include the anharmonic corrections 
perturbatively. Up to second-order in perturbation theory, we get the more
accurate expression
\beq
E_n ~=~ \sqrt{2 n q \delta} ~\bigl[ ~1 ~-~ \frac{n}{8} \frac{q}{\delta} ~-~
\frac{5n^2 +2}{128} \frac{q^2}{\delta^2} ~\Bigr] ~.
\label{en2}
\eeq
We can now include tunneling between wells; a WKB analysis shows that each
of the above energies should split into a band whose width $\Delta E_n$ is of 
the order of $\delta$ times $\exp (-2\delta /q)$. This explains why the bands 
rapidly become thin and shrink to isolated points as $q \rightarrow 0$ in Fig. 
2. In that figure, we have shown the six curves corresponding to $n=1$ to $6$ 
in Eq. (\ref{en2}); for $n=0$, we simply get a straight line lying at zero 
energy. We see that all the curves agree extremely well with the numerical 
data, even up to $E_n$ of the order of $\delta$ where the harmonic 
approximation breaks down and the band widths become noticeable.

We will now prove that the number of states lying below the line $\ed =1$ 
is proportional to $\delta$. For large $\delta /q$, we can use a 
semiclassical phase space argument to count how many states lie below any
given energy. The ``Hamiltonian" on the left hand side of Eq. (\ref{eom3}) 
has the form $p^2 + V(x)$, where $V(x) = \delta^2 \cos^2 (qx) - \delta q 
\sin (qx)$. Hence the number of states up to energy $E$ is given by the 
phase space integral
\beq
\nu (E) ~=~ \frac{1}{L} ~\int \int \frac{dx dp}{2\pi} ~\Theta (~ E^2 - p^2 - 
V(x) ~) ~, 
\label{phsp}
\eeq
where we have divided by the length $L$ of the system for normalization;
the $\Theta$-function is defined to be $1$ and $0$ if its argument is
positive and negative, respectively. On doing the momentum integral in 
(\ref{phsp}), we get
\beq
\nu (E) ~=~ \frac{q}{2\pi} ~\int_0^{2\pi /q} ~dx ~\frac{1}{\pi} ~\sqrt{E^2 - 
V(x)} ~\Theta (~E^2 - V(x) ~) ~,
\label{int}
\eeq
where we have used the fact that $V(x)$ has period $2\pi /q$. For small values 
of $\qd$, we then see that the number of states lying between $E=0$ 
and $\delta$ is 
\beq
\nu (\delta ) ~=~ \frac{2\delta}{\pi^2} ~,
\label{num1}
\eeq
which is {\it independent} of $q$. Incidentally, this also yields the number 
of bands lying below $E=\delta$. Since each band contains $q/\pi$ states, 
Eq. (\ref{num1}) gives the number of bands to be $2\delta/\pi q$. We have 
verified that this estimate agrees very well with our numerical results.

The difference in the spectrum for $q=0$ and $q \rightarrow 0$ should show
up most clearly in the specific heat at temperatures $T \ll \delta$. (We set
the Boltzmann constant equal to $1$ for convenience). For $q=0$, the
energy gap from $-\delta$ to $+\delta$ means that the specific heat vanishes
exponentially at low temperature. But for $q \rightarrow 0$, the semiclassical
expression in (\ref{int}) shows that the 
density of states at any energy $E \ll \delta$ is of the order of $\rho (E) 
\sim E/\delta$. This implies that the low-temperature specific heat goes as 
$T^2 /\delta$ which is very different from the exponential behavior at $q=0$.

Let us pause to ask: why does the model in Eq. (\ref{ham1}) show such 
different behaviors in the limit 
$q \rightarrow 0$ and at $q=0$? To answer this question, it is useful to
generalize the form of the incommensurate hopping in (\ref{ham1})
from $\delta \cos (\pi + q) n$ to $\delta \cos [(\pi + q)n +\eta ]$, where
$\eta$ is a phase lying between $0$ and $2\pi$. If $q/\pi$ is
{\it irrational} (this is the generic case to consider if $q$ is nonzero),
the spectrum does not depend on $\eta$. [If $q/\pi =M/N$ is rational, the 
spectrum does depend on $\eta$, but it varies less and less with $\eta$ as $N 
\rightarrow \infty$. In our numerical computations, we introduced a random 
phase $\eta$ for each value of $q$ and found that this had no noticeable 
effect on the two figures]. This insensitivity to $\eta$ can also be seen from 
the continuum theory in Eq. (\ref{eom1}) where a potential of the form 
$\cos (qx +\eta )$ can be transformed to $\cos (qx)$ by shifting $x$
appropriately. However, {\it exactly} at $q=0$, the hopping has a term like
$\delta \cos \eta \cos (\pi n)$ and the spectrum depends significantly on 
$\eta$; for instance, there is a gap from $E=0$ up to $\vert \delta \cos \eta 
\vert$, and the number of states up to an energy $E$ is given by
\beq
\nu (E, \eta ) ~=~ \frac{1}{\pi} ~\sqrt{E^2 ~-~ \delta^2 \cos^2 \eta} ~
\label{num2}
\eeq
for $\vert \delta \cos \eta \vert < E \ll 1$. We may now consider taking an 
{\it average} of this number over all possible values of $\eta$. Thus, the 
number of states up energy $E$ is given by Eq. (\ref{num2}) to be
\beq
\nu (E) ~=~ \int_0^{2\pi} ~\frac{d\eta}{2\pi^2} ~\sqrt{E^2 - \delta^2 \cos^2 
\eta} ~\Theta (~ E^2 - \delta^2 \cos^2 \eta ~)~.
\eeq
This agrees with (\ref{int}) if we substitute $\eta =qx$. Thus, 
the limit $q \rightarrow 0$ agrees
with the point $q=0$ provided that we average over $\eta$ in the latter case. 
However, this statement may not have any physical significance since an
experimental system sitting at $q=0$ (the dimerized point) is more likely 
to choose a particular value of $\eta$, rather than average over many values 
of $\eta$. We will therefore choose $\eta =0$ henceforth.

We will now argue that the difference between small $q$ and $q=0$ (in
particular, the presence of states within the $q=0$ gap) persists for the
more interesting case of interacting fermions, i.e., for Luttinger liquids.
To this end, consider adding a four-fermion interaction
term like $\sum_n c_n^\dagger c_n c_{n+1}^\dagger c_{n+1}$ to the
Hamiltonian (\ref{ham1}). Equivalently, we can consider an $XXZ$ spin-$1/2$
chain governed by
\beq
H ~=~ \frac{1}{2} ~\sum_n ~\Bigl[ ~( ~1 ~+~ \delta \cos ~(\pi + q)n ~) ~
(~S_n^x S_{n+1}^x ~+~ S_n^y S_{n+1}^y ~) ~+~ \Delta ~S_n^z S_{n+1}^z ~\Bigr] ~.
\label{ham2}
\eeq
which is related to the interacting fermion theory by a Jordan-Wigner
transformation \cite{lie}. If the incommensurate term is absent ($\delta =0$), 
it is known from the exact Bethe ansatz solution and conformal field theory
\cite{sch} that the model in (\ref{ham2}) is gapless at zero magnetic field if 
$-1 < \Delta \le 1$. If we now add a dimerization ($q=0$ and $\delta$ is
small), then the system becomes gapped and the gap scales as
\bea
\Delta E ~&=&~ \delta^{1/(2-K)} ~, \nonumber \\
{\rm where} \quad K ~&=&~ \frac{\pi}{2 \cos^{-1} (-\Delta)} ~.
\label{gap}
\eea
[Actually, a gap opens up only if $K \le 2$. If $K > 2$, the dimerization term
is irrelevant in the sense of the renormalization group, and a gap is not
generated]. It is useful to state the result in Eq. (\ref{gap}) in the 
language of bosonization. For $q=0$, the model in (\ref{ham2}) is equivalent, 
at low energies, to a bosonic sine-Gordon theory 
described by the Lagrangian density \cite{sch}
\beq
{\cal L} ~=~ \frac{1}{2\pi K} ~\Bigl[ ~(\partial \phi /\partial t)^2 ~-~
(\partial \phi /\partial x)^2  ~\Bigr] ~-~ \alpha ~\delta^{2/(2-K)} ~[~1 ~-~
\cos (2 \phi) ~] ~,
\label{lag1}
\eeq
where $\alpha$ is a positive constant whose numerical value is not important
here; the main point is to note the exponent of $\delta$ in the coefficient
of $\cos (2\phi)$. (There are also factors of $\ln \delta$ due to the presence
of a marginal operator in the $XXZ$ model \cite{aff}, but we will ignore 
such terms here). Thus, the incommensurate term which is {\it linear} in 
$\delta$ in the original theory in (\ref{ham2}) becomes a cosine
term with a different exponent for $\delta$ in the bosonic theory. This is 
because a nontrivial renormalization occurs in the process of deriving the 
low-energy bosonic theory from the fermionic one \cite{bha,wie}. This
renormalization occurs even if the fermions are noninteracting, i.e.,
for the $XY$ spin-$1/2$ chain with $\Delta =0$ in (\ref{ham2}) and $K=1$; the
reason for this is that the sine-Gordon theory is always strongly interacting. 
These strong interactions are also responsible for the large renormalizations 
of the correct quantum spectrum compared to the naive (i.e. classical)
spectrum that one obtains from the sine-Gordon theory, 
namely, the classical soliton mass for the fermionic excitations and the 
quadratic fluctuations around $\phi =0$ spectrum for the bosonic excitations
\cite{raj}. The noninteracting fermionic model in (\ref{ham1}) has no such 
renormalizations, which is why we did not use bosonization in the earlier
part of this paper.

Let us now change $q$ slightly away from $0$. Then bosonization 
again yields a theory of the sine-Gordon type, except that the coefficient of 
the $\cos (2\phi )$ term in Eq. (\ref{lag1}) gets modified to produce
\beq
{\cal L} ~=~ \frac{1}{2\pi K} ~\Bigl[ ~(\partial \phi /\partial t)^2 ~-~
(\partial \phi /\partial x)^2  ~\Bigr] ~-~ \alpha ~\vert \delta \sin (qx) 
\vert^{2/(2-K)} ~[~1 ~-~ \cos (2 \phi) ~] ~,
\label{lag2}
\eeq
where we have again shifted $x$ to convert $\cos (qx)$ to $\sin (qx)$. Unlike
(\ref{lag1}), the theory in Eq. (\ref{lag2}) cannot be solved analytically, 
either in classical or in quantum mechanics. However we can make some 
qualitative statements about the low-energy spectrum if $\qd \ll 1$. When
calculating the spectrum of small oscillations about a classical ground state
$\phi =0$, we find it convenient to scale $x$ by a factor of
\beq
a ~=~ (q \delta )^{1/(3-K)} ~.
\eeq
This gives the eigenvalue equation
\beq
-~\frac{\partial^2 \phi}{\partial x^2} ~+~ 4 \pi K \alpha ~\vert x 
\vert^{2/(2-K)} ~ \phi ~=~ \Bigl( \frac{E}{a} \Bigr)^2 ~\phi 
\label{schr}
\eeq
for eigenstates lying close to the origin $x=0$; we have approximated $\sin
(qx)$ by $qx$ and $\sin (2\phi )$ by $2\phi$. Since (\ref{schr}) is the
Schr\"odinger equation with a confining potential, we see that the energy 
can take several discrete values which are given by 
numerical factors multiplying $a$. [Note that our
earlier results for noninteracting fermions agree with this scaling argument 
if we set $K=1$]. These discrete values will then spread out into bands
once we include tunneling between the different wells centered at the
points $x=n \pi$.
Similarly, we can find the energy of a classical soliton which goes from
$\phi =0$ at $x \rightarrow - \infty$ to $\phi = \pi$ at $x \rightarrow 
\infty$; once again we can show by scaling that this will be given by $a$ times 
some numerical factor. [Of course, all the numerical factors will get
renormalized due to quantum corrections, but the power-law dependence on 
$q\delta$ is not expected to change]. We therefore see that all the low-lying 
excitations have energies of the order of $(q \delta)^{1/(3-K)}$; this is 
much smaller than the gap which exists exactly at $q=0$ given by $\Delta E$ in 
Eq. (\ref{gap}), since we are assuming that $\qd$ is small. We thus see that 
in the limit $q \rightarrow 0$, there are 
many states which lie within the gap $\Delta E$. 

We can use semiclassical arguments as in Eq. (\ref{int}) to estimate 
the density of states $\rho (E)$ at energies much smaller than the $q=0$ gap.
If $E \ll \delta^{1/(2-K)}$, the wave function for that state lies in a 
region where we can approximate $\sin (qx)$ by $qx$. Ignoring various 
numerical factors, we then get
\beq
\rho (E) ~=~ \frac{d}{dE} ~\int_0^{2\pi /q} ~\frac{q dx}{\pi} ~\sqrt{E^2 - 
(\delta q x)^{2/(2-K)}} ~\Theta (~E^2 - (\delta q x)^{2/(2-K)} ~) ~\sim ~
\frac{E^{2-K}}{\delta} ~.
\eeq
The specific heat at temperatures much lower than the $q=0$ gap therefore
scales as $T^{3-K} /\delta$, which is again much larger than the exponential 
dependence which occurs at $q=0$.

Before ending this paper, we should remark that the peculiar difference 
between the limit $q \rightarrow 0$ and $q=0$ only occurs near dimerization,
i.e., near wave number $\pi$. If we take the incommensurate term in Eq. 
(\ref{ham1})
to be of the form $\delta \cos (Q + q)n$, where $Q/\pi$ is equal to some
simple rational number {\it different} from $1$, then there is no such
singularity at $q=0$. Let us see this for the case of noninteracting fermions.
Imagine filling up the Fermi sea up to a Fermi energy $E_F = -\cos k_F$ where 
$k_F = Q/2$; the Fermi velocity is then $v_F = \sin k_F$. We define the
continuum Dirac field as 
\beq
\psi (x) ~=~ \psr (x) ~\exp ~( i k_F x) ~+~ \psl (x) ~\exp ~( - i k_F x) ~,
\eeq
After dropping terms like $\exp (\pm i k_F x)$ which vary rapidly, the Dirac 
equations take the form
\bea
i ~(~\partial /\partial t ~+~ v_F \partial /\partial x ~) ~\psr ~+~ 
\frac{\delta}{2} ~\exp ~(iqx-ik_F) ~\psl ~&=&~ 0 ~, \nonumber \\
i ~(~\partial /\partial t ~-~ v_F \partial /\partial x ~) ~\psl ~+~ 
\frac{\delta}{ 2} ~\exp ~(-iqx+ik_F) ~\psr ~&=&~ 0 ~. 
\eea
The solutions of this equation have the plane wave form
\beq
\Bigl( \begin{array}{c} 
\psr (x,t) \\
\psl (x,t)
\end{array} \Bigr) ~=~ \exp ~(-iEt) ~\Bigl( \begin{array}{c} 
a_R ~\exp ~(ikx + iqx/2 - ik_F/2) \\
a_L ~\exp ~(ikx - iqx/2 + ik_F/2) 
\end{array} \Bigr) ~.
\eeq
From this, we find that there is a single energy gap 
extending from $E=(v_F q-\delta )/2$ to $E=(v_F q+\delta )/2$; the 
size of the gap is independent of $q$, assuming that $q$ and $\delta$ are 
both small. If $\delta$ is held fixed and $q$ is increased from zero, a state 
first appears at zero energy when $q$ reaches the value $\delta /v_F$. 
It is instructive to express all this in the language of bosonization. Since
$K=1$, the sine-Gordon Lagrangian takes the form
\beq
{\cal L} ~=~ \frac{1}{2\pi v_F} ~(\partial \phi /\partial t)^2 ~-~ \frac{v_F}{
2\pi} (\partial \phi /\partial x)^2  ~-~ \frac{\pi \delta^2}{16 v_F} ~[~1 ~-~
\cos (2 \phi + qx) ~] ~,
\eeq
where we have fixed the coefficient of $\cos (2\phi )$ in such a way that,
for $q=0$, the soliton mass, including the quantum corrections \cite{raj}, 
is exactly equal to the gap from zero energy, i.e., $\delta /2$. It is 
convenient to scale $x$ in the action ($S =\int dt dx {\cal L}$) by a 
factor of $v_F$, and then shift $\phi$ by $q v_F x/2$. This gives
\beq
{\cal L} ~=~ \frac{1}{2\pi} ~\Bigl[ ~(\partial \phi /\partial t)^2 ~-~ 
(\partial \phi /\partial x ~-~ qv_F /2 )^2  ~\Bigr] ~-~ \frac{\pi 
\delta^2}{16} ~[~1 ~-~ \cos (2 \phi ) ~] ~.
\eeq
The Hamiltonian therefore contains a boundary term $-(qv_F /2\pi )
[\phi (\infty ) - \phi (-\infty )]$; this is equal to $-qv_F /2$ in a 
one-soliton state. We thus see that if $q$ is increased from zero, the 
energy of the one-soliton state becomes equal to that of the ground
state (whose soliton number is zero) when $q$ reaches the value $\delta /v_F$.
[We note that similar results appear in the context of incommensurate 
crystals \cite{lyu,bha,nak}]. 

Thus, if $Q \ne \pi$, there is a critical and nonzero value of $q$ at which 
the gap closes at zero energy. This is very
different from the earlier case with $Q=\pi$, where there is always 
a band of energies lying around zero energy, so that there is no
gap at zero energy for any $q \ne 0$. (Our results disagree with Ref. 7 which 
argues that there is a nonzero critical value of $q$ for $Q=\pi$).

An interesting problem for future study may be to examine what happens if
the incommensuration is strong, i.e., if $\delta$ is comparable to or larger
than $1$. [Both the perturbative and the continuum Dirac theory (or 
bosonization) approaches would then fail]. It is possible that the pattern
of bands and gaps would become much more complicated in that case, perhaps
with a Devil's staircase or a point spectrum structure \cite{sok,luc,ost,lyu}.
The nature of the wave functions, namely, whether they remain extended or
become localized, would also be of interest.

\newpage

\vskip .7 true cm
\noindent {\bf Acknowledgments}

We thank G. Ananthakrishna, Somen Bhattacharjee (who introduced us to this
problem), Rahul Pandit and, specially, Apoorva Patel for many useful 
discussions.

\vskip .5 true cm

\vskip 2 true cm
\noindent {\bf Figure Captions}
\vskip .5 true cm

\noindent
{1.} The bands as a function of the energy $E$ and the wave number $q$, both 
in units of $\delta$, taking $\delta =0.05$. The numbers $1$, $3$ and $5$ 
labeling the three biggest gaps are explained in the text.

\noindent
{2.} A finer view of the bands as a function of $E/ \delta$ and $q/ \delta$,
for $\delta = 0.05$. A comparison with the second-order perturbation 
results is indicated by the six solid lines.

\newpage

\begin{figure}[ht]
\begin{center}
\epsfig{figure=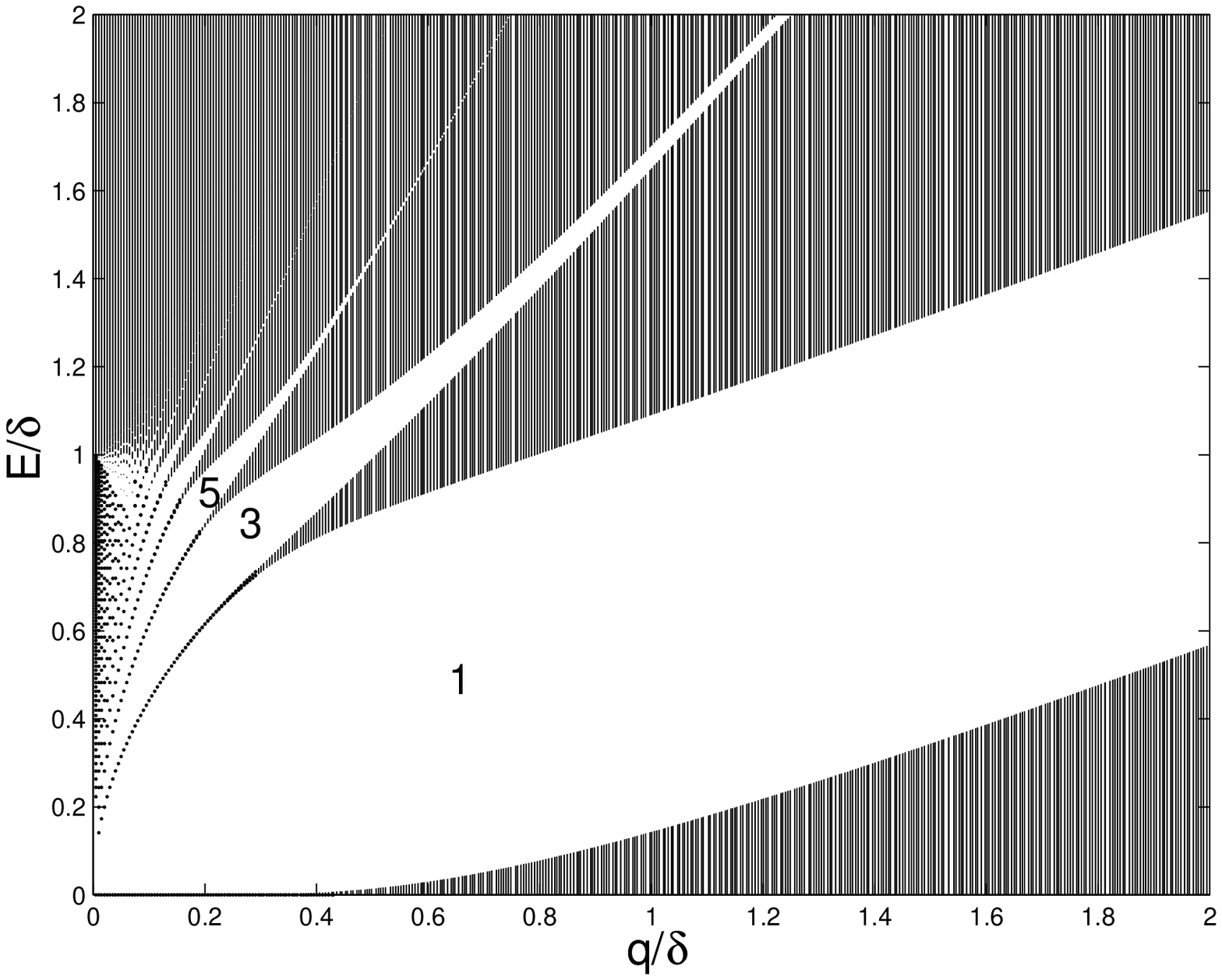,width=11cm}
\end{center}
\caption{}
\label{coarse}
\end{figure}

\vspace*{1cm}
\begin{figure}[hp]
\begin{center}
\epsfig{figure=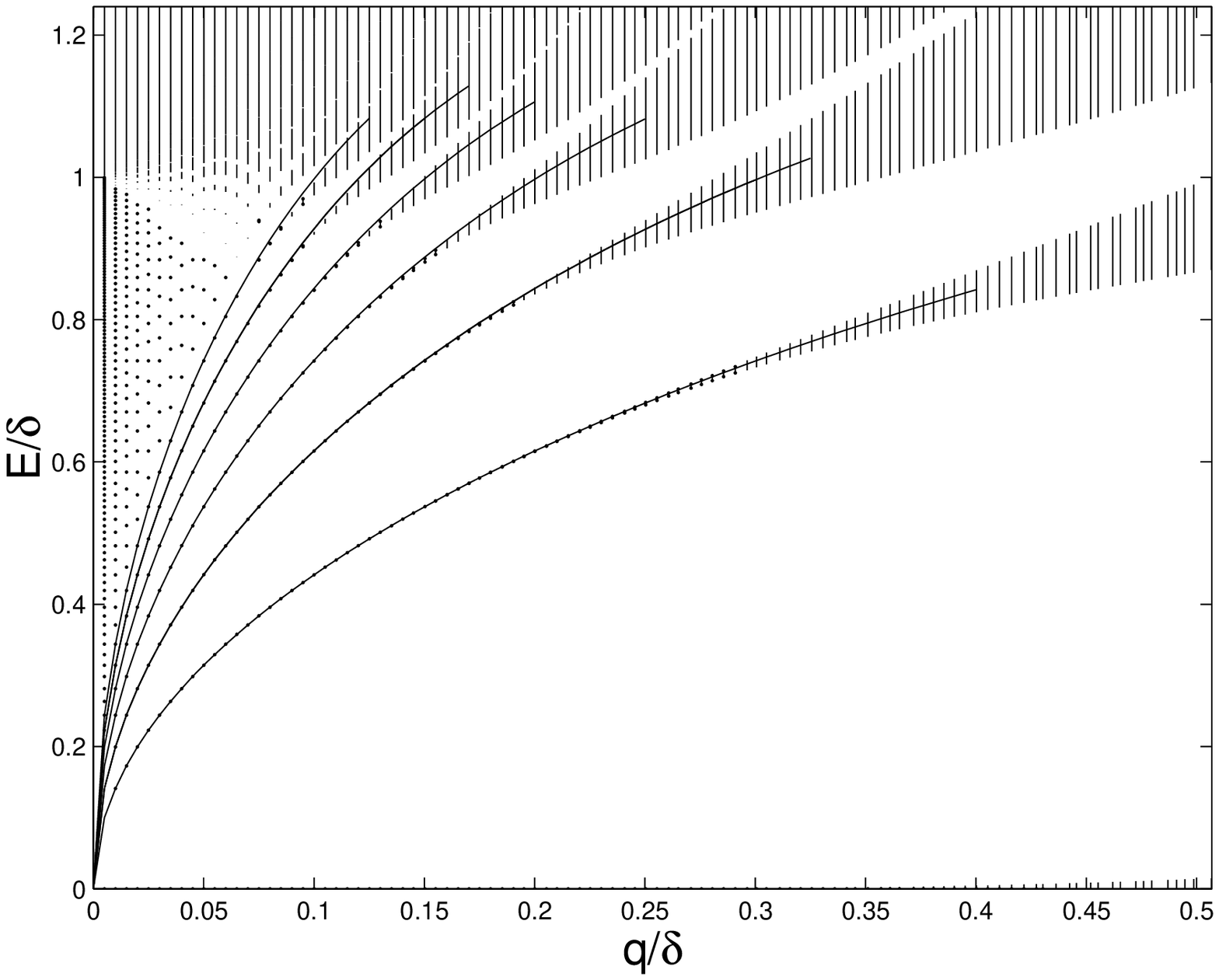,width=11cm}
\end{center}
\caption{}
\label{fine}
\end{figure}


\begin{thebibliography}{99}

\bibitem{sok} J. B. Sokoloff, Phys. Rep. {\bf 126}, 189 (1985).

\bibitem{luc} J. M. Luck, J. Stat. Phys. {\bf 72}, 417 (1993) and Phys. Rev.
B {\bf 39}, 5834 (1989).

\bibitem{val} M. C. Valsakumar and G. Ananthakrishna, J. Phys. C {\bf 20}, 9
(1987).

\bibitem{ost} S. Ostlund and R. Pandit, Phys. Rev. B {\bf 29}, 1394 (1984).

\bibitem{lyu} I. Lyuksyutov, A. G. Naumovets and V. Pokrovsky, {\it
Two-Dimensional Crystals} (Academic Press, San Diego, 1992);  P. M. Chaikin 
and T. C. Lubensky, {\it Principles of condensed matter physics} (Cambridge 
University Press, Cambridge, 1998).

\bibitem{mer} R. Merlin, K. Bajema, R. Clarke, F.-Y. Juang and P. K.
Bhattacharya, Phys. Rev. Lett. {\bf 55}, 1768 (1985).

\bibitem{bha} S. M. Bhattacharjee and S. Mukherji, J. Phys. A {\bf 31}, 
L695 (1998).

\bibitem{nak} T. Nakano and H. Fukuyama, J. Phys. Soc. Jpn. {\bf 49}, 1679
(1980); {\it ibid.} {\bf 50}, 2489 (1981).

\bibitem{abr} M. Abramowitz and I. A. Stegun, {\it Handbook of Mathematical
Functions} (Dover Publications, New York, 1972).

\bibitem{lie} E. Lieb, T. Schultz and D. Mattis, Ann. Phys. (N. Y.) {\bf 16},
407 (1961); P. Pfeuty, {\it ibid.} {\bf 57}, 79 (1970).

\bibitem{sch} H. J. Schulz, in {\it Strongly Correlated Magnetic and 
Superconducting Systems}, edited by G. Sierra and M. A. Martin-Delgado,
Lecture Notes in Physics 478 (Springer, Berlin, 1997); I. Affleck, in {\it 
Fields, Strings and Critical Phenomena}, edited by E. Brezin and J. 
Zinn-Justin (North-Holland, Amsterdam, 1989).

\bibitem{aff} I. Affleck, D. Gepner, H. J. Schulz and T. Ziman, J. Phys. A
{\bf 22}, 511 (1989).

\bibitem{wie} P. B. Wiegmann, J. Phys. C {\bf 11}, 1583 (1978).

\bibitem{raj} R. Rajaraman, {\it Solitons and Instantons} (North-Holland, 
Amsterdam, 1982).

\end{thebibliography}
\end{document}